\documentclass[journal, final]{IEEEtran}

\usepackage{tcolorbox}
\newtcolorbox{mybox}{colback=blue!1!white,colframe=blue!75!black}

\usepackage[linesnumbered,lined, algoruled]{algorithm2e}
\usepackage{amssymb,bm}
\usepackage{amsmath}
\usepackage[T1]{fontenc}
\usepackage[acronym,shortcuts]{glossaries}
\usepackage{algorithmic,float}
\usepackage{setspace, url}
\usepackage{balance}
\usepackage{color}
\usepackage{graphicx}
\usepackage{graphics}
\usepackage{siunitx}
\usepackage{subfigure}
\usepackage{comment}
\usepackage{dsfont}
\usepackage{soul}
\usepackage{cite}
\usepackage{booktabs}
\usepackage{amsthm}

\newacronym{3gpp}{3GPP}{$3^{\rm rd}$ generation partnership project}
\newacronym{5g}{b5G}{beyond the fifth generation}
\newacronym{aod}{AOD}{angle of departure}
\newacronym{amise}{AMISE}{asymptotic mean integrated squared error}
\newacronym{awgn}{AWGN}{additive white Gaussian noise}
\newacronym{bler}{FT-BER}{first transmission block error rate}
\newacronym{cdf}{CDF}{cumulative density function}
\newacronym{dct}{DCT}{discrete cosine transform}
\newacronym{gnb}{gNB}{next-generation node base}
\newacronym{iid}{i.i.d}{independent identically distributed}
\newacronym{ipla}{IPLA}{interference-prediction LA}
\newacronym{ipv}{IP}{interference power}
\newacronym{kde}{KDE}{kernel density estimator}
\newacronym{la}{LA}{link adaptation}
\newacronym{lcsb}{LC-SB}{low-complexity \ac{sbsse}}
\newacronym{los}{LOS}{line of sight}
\newacronym{mcs}{MCS}{modulation and coding scheme}
\newacronym{mq}{MQ}{maximum quantile}
\newacronym{mrt}{MRT}{maximal ratio transmission}
\newacronym{olla}{OLLA}{outer loop link adaptation}
\newacronym{pdf}{p.d.f.}{probability density function}
\newacronym{rr}{RR}{round robin}
\newacronym{rv}{r.v.}{random variable}
\newacronym{sbsse}{SB-KDE}{subsets-based \ac{kde}}
\newacronym{se}{DR}{data rate}
\newacronym{sinr}{SINR}{signal to interference plus noise ratio}
\newacronym{snr}{SNR}{signal to noise ratio}
\newacronym{tti}{TTI}{transmission time interval}
\newacronym{ue}{UE}{user equipment}
\newacronym{ula}{ULA}{uniform linear array}
\newacronym{urllc}{URLLC}{ultra reliable low latency communications}
\newacronym{vbkde}{VB-KDE}{variable bandwidth \ac{kde}}

\newtheorem{theorem}{Theorem}
\newtheorem{lemma}[theorem]{Lemma}

\title{Interference Prediction for Low-Complexity Link Adaptation in Beyond 5G Ultra-Reliable Low-Latency Communications}
\author{Alessandro Brighente$^*$, Jafar Mohammadi$^\dagger$, \\
Paolo Baracca$^\dagger$, Silvio Mandelli$^\dagger$,
and Stefano Tomasin$^*$\\
$^*$Department of Information Engineering, University of Padova, Italy\\
$^\dagger$Nokia Bell Labs, Nokia, Germany }

\begin{document}

\maketitle

\sloppy
\begin{abstract}
Traditional \ac{la} schemes in cellular network must be revised for  networks \ac{5g}, to guarantee the strict latency and reliability requirements advocated by \ac{urllc}. In particular, a poor error rate prediction potentially increases retransmissions, which in turn increase latency and reduce reliability. In this paper, we present an interference prediction method to enhance \ac{la} for \ac{urllc}. To develop our prediction method, we propose a kernel based probability density estimation algorithm, and provide an in depth analysis of its statistical performance. We also provide a low complxity version, suitable for practical scenarios. The proposed scheme is compared with state-of-the-art \ac{la} solutions over fully compliant \ac{3gpp} calibrated channels, showing the validity of our proposal. 
\end{abstract}
\begin{IEEEkeywords}
Beyond 5G, kernel distribution estimation, statistical link adaptation, ultra-reliable low-latency communications. 
\end{IEEEkeywords}
\glsresetall
\section{Introduction}
Among the different application scenarios for networks \ac{5g}, \ac{urllc} have drawn significant attention from both industrial and academic research. The strict requirements on latency ($1-10$ ms) and reliability (\ac{bler} $< 10^{-5}$) will enable new use cases, such as factory automation \cite{ren2019joint}, autonomous vehicles \cite{ge2019ultra}, and tactile Internet \cite{kim2018ultrareliable}.

In order to meet the aforementioned requirements, different communication solutions have been investigated. In fact, short packets \cite{shirvanimoghaddam2018short}, shorter transmission time intervals, and grant-free access schemes \cite{zhou2018resource} are  enablers of \ac{urllc}. The reader is referred to \cite{li20185g} for an overview of available solutions.

In this paper, we focus on \ac{la}, i.e., the choice of a proper modulation and coding scheme such that a certain \ac{bler} is met. \ac{la} drastically reduces the number of required retransmissions, as the choice of the \ac{mcs} is based on the channel quality at transmission time, therefore adapting transmissions to the actual channel quality. However, in order to meet very low target \ac{bler}, a poor \ac{la} may yield too conservative transmission rates, with waste of resources. Instead, the  \ac{la} algorithm should  guarantee the target \ac{bler}, while at the same time avoiding too conservative behaviors. 

Different solutions have been proposed for \ac{la} in \ac{urllc} context. 
In \cite{pocovi2018} \ac{la} is performed based on a filtered version of the \ac{sinr}, where the \ac{ipv} for the next transmission is predicted by low-pass filtering past \acp{ipv}. Similarly, in \cite{pocovi2020channel}, prediction is obtained by two different low-pass filters,  designed  on the difference between  current and  previous filtered \ac{ipv}. However, this approach is sensitive to high oscillations of \acp{ipv},  yielding a  sub-optimal \ac{la}.
In \cite{shariatmadari2016link}, a \ac{la} solution has been proposed to attain ultra-reliability. However, this is obtained by means or retransmission, which increase the overall latency. When delay constraints allow for retransmissions, proper schemes can achieve the desired reliability with higher spectral efficiency \cite{buccheri2018hybrid}. When delay constraints becomes strict, instead, the \ac{bler} at the first transmission becomes the key enabler for \ac{urllc}. In \cite{belogaev2019conservative} a conservative \ac{la} algorithm has been proposed, where the \ac{mcs} is chosen on the basis of the estimated  strongest channel degradation at the packet transmission time. Although retransmissions may not be needed using a conservative \ac{mcs}, this solution does not  fully exploit  channel conditions at transmission time. In \cite{deghel2018joint}, a joint \ac{la} and retransmission policy is obtained, based on the average \ac{sinr} value, which however cannot guarantee  a small \ac{bler} in the short term.
In \cite{vtc}, \ac{la} is implemented based on \acp{ipv} statistics. In particular,  the \ac{pdf} of  \ac{ipv} is predicted using \ac{kde} \cite{parzen1962}, and then used in the \ac{la} algorithm. 

In this paper, following the approach in \cite{vtc}, we propose an algorithm for the estimation of the \ac{ipv} \ac{pdf} and its use for \ac{la}. The proposed solution is evaluated in a cellular network scenario with channels following either the  Rice channel model or a  \ac{3gpp} calibrated  3D Urban micro (UMi) model \cite{138901}. It turns out that the proposed \ac{la} framework is more accurate and entails a lower complexity than state-of-the-art solutions. 

With respect to the literature, and in particular to \cite{vtc}, the contribution of this paper are the following:
\begin{itemize}
	\item we propose a novel density estimation algorithm, i.e., \ac{sbsse}, and compare its performance with the state-of-the-art \ac{kde};
	\item we propose a low-complexity version of \ac{sbsse}, i.e., \ac{lcsb};
	\item we show how density estimation methods can be leveraged for \ac{la}, and assess their performance, comparing it with \ac{kde} (for \ac{pdf} prediction), \ac{olla} and a log-normal approximation for \ac{la};
	\item we test our solution in a \ac{3gpp} calibrated 3D UMi scenario, further assessing the validity of the proposed framework.
\end{itemize}

The rest of the paper is organized as follows. In Section~\ref{sec:sysMod} we introduce the system model for the considered cellular network, and we review \ac{la}. In Section~\ref{sec:intPred} we present interference prediction for \ac{la}, and review \ac{kde}. In Section~\ref{sec:sbsse} we introduce \ac{sbsse} and its low-complexity version. The computational complexity analysis of all algorithms is derived in Section~\ref{sec:compcomp}. In Section~\ref{sec:numRes} we assess the validity of the proposed framework, by comparing it with state-of-the-art algorithms in both a Rice channel model and a \ac{3gpp} calibrated 3D Umi scenario. Lastly, in Section~\ref{sec:conclusion} we draw the conclusions.

\section{System Model}\label{sec:sysMod}

We consider a cellular network with $C$ cells wherein, for each cell, a single \ac{gnb} equipped with $N_a$ antennas serves $K_{\rm tot}$ single-antenna \acp{ue}. Each cell is populated by a random number of \acp{ue} uniformly located in space. We denote as $\mathcal{U}_c$ the set of \acp{ue} indexes in cell $c$. We assume that each \ac{gnb} serves a single \ac{ue} in a resource block, and we denote as $\bm{h}(c,t)$ the $\mathbb{C}^{1 \times N_a}$ channel from \ac{gnb} $c$ toward the \ac{ue} served at \ac{tti} $t$, in the considered resource block. 



We consider two different channel models. We first consider the Rice channel model \cite[Ch. 2.4.2]{tse2005}, where each link is a linear combination of a \ac{los} and non \ac{los} links. While this model is widely used in the literature for its implementation simplicity, we also consider a more realistic spatial 3D UMi \ac{3gpp} calibrated \cite{138901} scenario.

In both cases, we assume that downlink transmissions are performed using the \ac{mrt} precoder 
\begin{equation}
\bm{g}(c,t) = \frac{\bm{h}^H(c,t)}{||\bm{h}(c,t)||},
\end{equation}
where $[\cdot]^H$ denotes the Hermitian of a vector.


The signal received by the served \ac{ue} suffers from the interference caused by all \acp{gnb} with index $\ell =1, \cdots, C, \, \ell \neq c$,  transmitting toward their scheduled \acp{ue}. The \ac{sinr} measured at \ac{ue} in cell $c$ at \ac{tti} $t$ is given by 
\begin{equation}\label{eq:sinr}
\rho(c,t) = \frac{|\bm{h}(c,t)\bm{g}(c,t)|^2P}{\phi(t)+ \sigma^2},
\end{equation}
where $P$ is the transmitted power, $\sigma^2$ is the noise power, $\phi(t)$ is the \ac{ipv} due to other scheduled users at \ac{tti} $t$, i.e.,
\begin{equation}
\phi(t) = \sum_{\ell = 1, \ell \neq c}^{C}|\bm{h}(\ell,t)\bm{g}(\ell,t)|^2P,
\end{equation}
and $x_\ell(t)$ denotes the index of the \ac{ue} served by the $\ell$-th \ac{gnb} at \ac{tti} $t$ .

We focus on the most extreme \ac{urllc} cases (e.g., motion control) that are characterized by deterministic periodic traffic \cite{22104} and assume that a) packets for a given \ac{ue} appear at periodic \acp{tti} and b) \acp{ue} are served by the \acp{gnb} in a deterministic fashion, according to a \ac{rr} scheduler that allows each packet to meet its latency constraints. The use of a \ac{rr} scheduler, implies that $\rho$ values are correlated in time, since periodically each user will be affected by the same subset of interferers. Correlation is exploited in the design of our proposed \ac{la} scheme. Then, without loss of generality, we assume that \acp{ue} are sequentially served according to their index in $\mathcal{U}_c, \, c=1,\ldots,C$. Moreover, because of the strict latency requirements of \ac{urllc} traffic, we assume that no retransmission is allowed. Therefore, packets that are successfully received at the \acp{ue} always meet the latency constraint, whereas, when  transmission fails, the packet is dropped. Finally, we assume that each \ac{gnb} is fully loaded, i.e., in each \ac{tti} there is always a certain \ac{ue} that needs to be scheduled by each \ac{gnb}. 

In this paper we focus on \ac{la}, i.e., on the problem of choosing a proper \ac{mcs}, subject to a first-transmission target \ac{bler}. Whilst for each \ac{mcs} index the values of modulation order, target code rate, and spectral efficiency are given (see, e.g., \cite{138214}), \ac{bler} values are usually obtained by simulations or data collection. For a given \ac{bler}, the minimum \ac{sinr} needed for each \ac{mcs} is stored in a look-up table. Since now we will focus on the \ac{la} of a single \ac{gnb}, we drop index $c$ from the notation. In particular, considering the set $\mathcal{M}$ of available \acp{mcs} and assuming that \acp{mcs} are ordered by their increasing rates, the \ac{la} problem for \ac{tti} $t+1$ can be written as
\begin{equation}\label{eq:laProb}
    {\rm M}^*(t+1) = \underset{i \in \mathcal{M}}{\max} \{{\rm M}_i : \hat{\rho}(t+1) > \rho_i \},
\end{equation}
where $\rho_i$ is the minimum \ac{sinr} for which the target \ac{bler} is achieved with \ac{mcs} $i$, and $\hat{\rho}(t+1)$ is an estimate of the \ac{sinr} at $t+1$. We see that the \ac{la} problem selects, among the  \acp{mcs} which guarantee a certain target \ac{bler}, the one which also maximizes the rate. 

The predicted \ac{sinr} in \ac{olla} \cite{sampath1997}  is obtained only from the last measured \ac{sinr} $\rho(t)$. The idea is that, if the last transmission was successful with the previously selected \ac{mcs} and an acknowledgment (ACK) packet is sent back to the receiver, the estimated \ac{sinr} $\rho_{\rm OLLA}(t+1)$ is increased. If instead the previous transmission failed and a non-acknowledgment (NACK) packet is sent back to the transmitter, and $\rho_{\rm OLLA}(t+1)$ is hence reduced. In particular, the  \ac{sinr} value used for \ac{mcs} selection at \ac{tti} $t+1$ is
\begin{equation}
    \hat{\rho}_{\rm OLLA}(t+1) = \hat{\rho}(t)+\Delta(t+1),\label{eq:olla}
\end{equation}
where the offset is computed as 
\begin{equation}
    \Delta(t+1) =
    \begin{cases}
    \Delta(t) + \Delta_{\rm ACK}, {\rm if ACK at  t}; \\
    \Delta(t) + \Delta_{\rm NACK}, {\rm if NACK at t}.
    \end{cases}
\end{equation}
The values $\Delta_{\rm ACK}>0$ and $\Delta_{\rm NACK}$ are suitably selected such that the target \ac{bler} is met, with $\Delta_{\rm NACK} = -\frac{1-\varepsilon}{\varepsilon}\Delta_{\rm ACK}$ \cite{sampath1997}.  Therefore, in \eqref{eq:olla} the \ac{sinr} estimate is typically reduced in order to have a more conservative approach in \ac{la}. 

We notice that the basic \ac{olla} design is not suitable for \ac{urllc} for two reasons. First, due to the very low \ac{bler} requirements of \ac{urllc}, adjusting the estimated \ac{sinr} via $\Delta_{\rm NACK}$ would lead to a  conservatively low \ac{mcs}. Indeed, to recover from a loss, \ac{olla} needs a number of steps proportional to the inverse of the required reliability, significantly  increasing  latency. The second drawback of \ac{olla} is that the target \ac{bler} is guaranteed over a window of duration $1/\varepsilon$, whereas the instantaneous level may be extremely different.

\section{Interference Prediction for LA}\label{sec:intPred}

In this paper, similarly to \cite{vtc}, we compute the predicted \ac{sinr} $\hat{\rho}(t+1)$ using the last $N_{\rm prev}$ measured $\rho(t)$, i.e., given the vector
\begin{equation}
\bm{\phi}_{N_{\rm prev}}(t)=[\phi(t),\cdots,\phi(t-N_{\rm prev}+1)].
\end{equation}
We also assume that we are perfectly able to track the channel of the scheduled user and we assume that only interference is rapidly changing  \cite{pocovi2018}. We aim at predicting the outage \ac{sinr}, i.e., an \ac{sinr} value that will be exceeded with high probability, thus reducing the probability of transmission failures. Thus, we resort to the \ac{mq} method proposed in \cite{vtc}, so that the \ac{sinr} prediction reduces to predicting the outage \ac{ipv} $\hat{\phi}(t+1)$, i.e., the value $\hat{\phi}(t+1)$ that satisfies
\begin{equation}
\mathbb{P}\left[ \hat{\phi}(t+1) < \phi(t+1)|\bm{\phi}_{N_{\rm prev}}(t)\right] \leq \varepsilon. 
\label{fondamentale}
\end{equation}
The resulting \ac{la} scheme is denoted as \ac{ipla}.
 
In order to compute $\hat{\phi}(t+1)$, we estimate the conditional \ac{pdf} of $\phi(t+1)$, given the observations $\bm{\phi}_{N_{\rm prev}}(t)$, i.e.,  $\hat{f}_{\phi(t+1)|\bm{\phi}_{N_{\rm prev}}(t)}(a | \bm{\phi})$. Then, \eqref{fondamentale} becomes 
\begin{equation}\label{eq:integrale}
\begin{split}
\mathbb{P}&\left[  \hat{\phi}(t+1) < \phi(t+1)|\bm{\phi}_{N_{\rm prev}}(t)\right] \\ & =\int_0^{\phi(t+1)}f_{\phi(t+1)|\bm{\phi}_{N_{\rm prev}}(t)}(a | \bm{\phi})  da 
\\ & \approx \int_0^{\phi(t+1)}\hat{f}_{\phi(t+1)|\bm{\phi}_{N_{\rm prev}}(t)}(a | \bm{\phi})  da .
\end{split}
\end{equation}
The integral in \eqref{eq:integrale} is computed via numerical integration.

Now, we still have the problem of obtaining an estimate of the conditional \ac{pdf} $\hat{f}_{\phi(t+1)|\bm{\phi}_{N_{\rm prev}}(t)}(a | \bm{\phi})$. To this end, we consider 
\begin{equation}
    L >> N_{\rm prev}
\end{equation}
samples of $\phi(t)$, i.e., $\phi(t-1), \ldots, \phi(t-L)$, which are used to estimate the conditional \ac{pdf}.  Then, we decompose the conditional \ac{pdf} into the ratio of the joint and marginal \acp{pdf} $f_{\phi(t+1),\bm{\phi}_{N_{\rm prev}}(t)}(a, \bm{\phi})$ and $f_{\bm{\phi}_{N_{\rm prev}}(t)}(\bm{\phi})$, i.e.,
\begin{equation}\label{eq:condPDF}
f_{\phi(t+1)|\bm{\phi}_{N_{\rm prev}}(t)}(a | \bm{\phi}) = \frac{f_{\phi(t+1),\bm{\phi}_{N_{\rm prev}}(t)}(a, \bm{\phi})}{f_{\bm{\phi}_{N_{\rm prev}}(t)}(\bm{\phi})}.
\end{equation}


In the following we will propose techniques to estimate the joint \ac{pdf} of multiple random variables, which will be used to estimate both the joint and marginal \acp{pdf} in \eqref{eq:condPDF}. In order to simplify notation, we will consider a generic random vector $\bm{x}$, with \ac{pdf} $f(\bm{x})$. The estimated \ac{pdf} will be obtained using  set  $\bm{\mathcal{S}} = \{\bm{s}_1, \cdots, \bm{s}_N\}$ of observed realizations of $\bm{x}$, where each element $\bm{s}_n$ is given by $N_{\rm prev}+1$ successive elements $\phi(\cdot)$, i.e., $\bm{s}_n = [\phi(t-2-n),\ldots, \phi(t-2-n - N_{\rm prev})]$. Notice that, since each $\bm{s}_n$ is obtained by grouping identically distributed elements, elements in $\bm{\mathcal{S}}$ are identically distributed.

\subsection{\ac{pdf} Estimation by Cumulative Density Function}\label{sec:kde}

The simplest \ac{pdf} estimator is obtained by means of histogram, i.e.,
\begin{equation}\label{eq:hist}
    \hat{f}_{\rm h}(\bm{x}) = \frac{\sum_{n=1}^N \delta(\bm{s}_n-\bm{x})}{N},
    \end{equation}
where $\delta(x)$ is the indicator function such that $\delta(x) = 1$ if $x=0$, $0$ otherwise.

Let us split the length $N_{\rm prev}+1$ vectors $\bm{x}$ and $\bm{s}_n$ into $\bm{x} = [a,\bar{\bm{x}}]$ and $\bm{s}_n = [b,\bar{\bm{s}}_n]$, such that $a$ and $b$ denote the first element, whereas $\bar{\cdot}$ denotes the remaining $N_{\rm prev}$ values. Based on \eqref{eq:hist}, we have the following
\begin{lemma}
The empirical conditional \ac{cdf} of $a$ given $\bar{\bm{x}}$ is given by  
\begin{equation}\label{eq:empCDF}
\hat{F}(a|\bar{\bm{x}})  = \frac{\sum_{n=1}^N\mathds{1}(b-a|\bar{\bm{x}})}{ \sum_{n=1}^N\delta(\bar{\bm{s}}_n-\bar{\bm{x}})}, 
\end{equation}
where $\mathds{1}({s}_n(1)-{x}(1)|\bar{\bm{x}}) = 1$ if $s_{n}(1) < x(1)$ given $\bar{\bm{s}}_n = \bar{\bm{x}}$, $0$ otherwise.
\begin{proof}
The result is given by substitution of \eqref{eq:hist} in \eqref{eq:condPDF}, and by applying the integration in \eqref{eq:integrale}.
\end{proof}
\end{lemma}
A drawback of this approach is that, when $L$ is not large enough, the entries with lower probability  will not appear in $\bm{\mathcal{S}}$, and the value of their \ac{pdf} will be zero. This problem becomes more prominent for \ac{urllc}, as targeting $10^{-5}$ or lower \ac{bler} requires a precise  \ac{pdf} estimate.

\subsection{\ac{kde}}

A more accurate \ac{pdf} estimator is  \ac{kde}, firstly introduced in \cite{parzen1962}, namely 
\begin{equation}\label{eq:parzen}
    \hat{f}\left(\bm{x}\right) = \frac{1}{N}\sum_{n=1}^N K\left(\frac{\bm{s}_n-\bm{x}}{h} \right),
\end{equation}
where $K$ is the \textit{kernel function} and $h$ is the \textit{kernel bandwidth}, i.e., a parameter of the kernel function which must be suitably selected.

We consider the multivariate Gaussian kernel with uncorrelated dimensions in \cite{botev2010}, i.e.,
\begin{equation}
    K\left(\frac{\bm{s}_n-\bm{x}}{h} \right) = \frac{1}{\sqrt{(2\pi)^D h}}\exp\left(-\frac{||\bm{s}_n-\bm{x})||^2}{2h} \right),
\end{equation}
where $D$ is the dimension. Bandwidth $h$ can be obtained by minimizing the \ac{amise}  \cite{botev2010}. In particular, when considering a Gaussian kernel, the \ac{amise} can be written as \cite{botev2010}
\begin{equation}\label{eq:amiseKde} 
{\rm AMISE}(h) = \frac{1}{4}\Upsilon(f^{''})h^4 + \frac{1}{2N\sqrt{\pi}h},
\end{equation}
where $f^{''}$ denotes the second derivative of the true \ac{pdf} $f$, and
\begin{equation}\label{eq:rDef}
    \Upsilon(f^{''}) = \int_{-\infty}^\infty [f^{''}(\bm{u})]^2 d\bm{u}.
\end{equation}
The minimum \ac{amise} value is obtained for the bandwidth
\begin{equation}\label{eq:optBWkde}
    h^* = \left(\frac{1}{2N\sqrt{\pi}\Upsilon(f^{''})} \right)^{\frac{1}{5}},
\end{equation}
which is the standard deviation of the Gaussian kernel. The assumption behind  \eqref{eq:parzen} is that values are independent identically distributed. We already discussed the latter condition when creating sets. However, due to the \ac{rr} scheduler, values in $\mathcal{\bm{S}}$ are correlated. This will be exploited to improve the quality of the proposed estimator, as detailed in Section~\ref{sec:lowComp}.

Notice that, by exploiting Gaussian kernels, a kernel-based estimate of the \ac{cdf} can be obtained by substituting the Gaussian function with the Gaussian Q function. However, we here focus on \ac{pdf} estimation.

\subsection{Variable-Bandwidth \ac{kde}}\label{sec:vb}
The choice of a single bandwidth value may not be accurate enough. For instance, it could be useful to have a larger bandwidth in intervals wherein few samples have been collected, and a smaller bandwidth where many samples has been collected.
In order to obtain a better estimate of the \ac{ipv}'s \ac{pdf}, we consider the \ac{vbkde}, i.e., a \ac{kde} with multiple bandiwdths. Among
\acp{vbkde} we find two different classes \cite{terrell1992}: the \textit{balloon estimator} and the \textit{sample smoothing estimator}. 

In the balloon estimator, the bandwidth $h$ is a function of the target point $\bm{x}$, i.e.,
\begin{equation}
    \hat{f}\left(\bm{x}\right) = \frac{1}{Nh(\bm{x})}\sum_{n=1}^N K\left(\frac{\bm{s}_n-\bm{x}}{h(\bm{x})} \right).
\end{equation}
In the sample smoothing estimator, the bandwidth $h$ is a function of the measured sample point $\bm{s}_n$, i.e.
\begin{equation}
    \hat{f}\left(\bm{x}\right) = \frac{1}{N}\sum_{n=1}^N \frac{1}{h(\bm{s}_n)}K\left(\frac{\bm{s}_n-\bm{x}}{h(\bm{s}_n)} \right),
\end{equation}
Both estimators present drawbacks: the balloon estimator has been showed to perform better than the fixed bandwidth \ac{kde} only when dealing with a multidimensional \ac{pdf} with more than three dimensions  \cite{terrell1992}. On the other hand, the sample smoothing estimator has been showed to be highly dependent on the distance between sample points. 

\section{Subsets-Based Sample Smoothing Estimator}\label{sec:sbsse}
In this paper, we focus on the second class of \ac{vbkde} estimators and propose a new method which deals with the drawback of sample smoothing estimators.

Let us consider the sample space (i.e., the list of possible outcomes) $\bm{\mathcal{X}}$ associated to the \ac{pdf} $f(\bm{x})$. We split $\bm{\mathcal{X}}$ in $B$ subsets, such that $\bm{\mathcal{X}}_i \cap \bm{\mathcal{X}}_j = \emptyset$, $\bigcup_{i=1}^B\bm{\mathcal{X}}_i = \bm{\mathcal{X}}$. Subsets are created according to a predefined rule, homogeneous in all dimensions, i.e., denoting as $\mathcal{X}^{(d)}=\{\mathcal{X}^{(d)}_1,\ldots,\mathcal{X}^{(d)}_B\}$ the group of subsets created along the $d^{\rm th}$ dimension, $\bm{\mathcal{X}}_i$ is the $i^{\rm th}$ elements of the cartesian product
\begin{equation}
\mathcal{X}^{(1)} \times \mathcal{X}^{(2)} \times \ldots \times \mathcal{X}^{(D)}.
\end{equation}
The rule deciding how subsets are created may be based on different factors, such as the value assumed by the elements in the set or the number of samples with close values. Both policies will be discussed in Section \ref{sec:la_nr}. 

Then, we consider a bandwidth value $h_i$ for each subsets, and define the \ac{sbsse} estimator
\begin{equation}\label{eq:sbkde}
    \tilde{f}(\bm{x}) = \frac{1}{B}\sum_{i=1}^B\frac{1}{|\bm{\mathcal{X}}_i|h_i}\sum_{\bm{s} \in \bm{\mathcal{X}}_i}K\left(\frac{\bm{s}-\bm{x}}{h_i} \right).
\end{equation}
Let $\mathbb{P}(\bm{\mathcal{X}}_{i})$ be the probability of the $i^{\rm th}$ subset, and $f(\bm{x}|\bm{\mathcal{X}}_{i})$ the \ac{pdf} of sample $\bm{x}$ in the $i^{\rm th}$ subset. Assuming that the value $B$ is given and that sets are created according to a certain policy, we have the following results on the \ac{sbsse}.
\begin{lemma}\label{lem:bias}
Given the number of subsets $B$, the bias of the \ac{sbsse} estimator is given by
\begin{IEEEeqnarray}{lCr}
  & {\rm Bias}(\tilde{f}(\bm{x})) \\
  & = \omega_2(K)\sum_{i=1}^B\frac{1}{2}f^{''}(\bm{x}|\bm{\mathcal{X}}_{i})h_{i}^2\mathbb{P}(\bm{\mathcal{X}}_{i}) + o(h_i^2), \nonumber
\end{IEEEeqnarray}
where $\omega_2(K)$ is the second moment of the kernel function $K$, i.e.,
\begin{equation}\label{eq:omegaDef}
\omega_2(K) = \int_{-\infty}^\infty \bm{u}^2 K(\bm{u}) d\bm{u}.
\end{equation}
\end{lemma}
\begin{proof}
See Appendix \ref{app:est}.
\end{proof}

\begin{lemma}\label{lem:var}
Given the number of subsets $B$, the variance of the \ac{sbsse} estimator is given by
\begin{IEEEeqnarray}{lCr}
  & {\rm Var}(\tilde{f}(\bm{x})) = \Upsilon(K)f(\bm{x})\frac{1}{B}\sum_{i=1}^B \frac{1}{h_i}  + O\left(\frac{1}{N}\right), \nonumber
\end{IEEEeqnarray}
where $N$ is the total number of samples.
\end{lemma}
\begin{proof}
See Appendix \ref{app:est}.
\end{proof}

Exploiting the results in Lemma \ref{lem:bias} and Lemma \ref{lem:var}, we also obtain the following two results.
\begin{theorem}
For a given number of subsets $B$, and bandwidths $\bm{h}=[h_1, \cdots, h_{B}]$, the \ac{amise} obtained via \ac{sbsse} exploiting a Gaussian kernel is given by
\begin{equation}\label{eq:amiseMult}
\begin{split}
& {\rm AMISE}(\bm{h}) = \\
& \frac{1}{4}\sum_{i=1}^B \left(\Upsilon(f^{''}(\bm{x}|\bm{\mathcal{X}}_i))h^4\mathbb{P}\left(\bm{\mathcal{X}}_i \right)^2 + \frac{1}{2B^2|\bm{\mathcal{X}}_i|\sqrt{\pi}h_i}\right).
\end{split}
\end{equation}
\end{theorem}
\begin{proof}
See Appendix \ref{app:est}.
\end{proof}

\begin{theorem}
For a given number of subsets $B$ and considering a Gaussian kernel, the minimum \ac{amise} is obtained via the \ac{sbsse} for the $i^{\rm th}$ bandwidth value given by
\begin{IEEEeqnarray}{lCr}\label{eq:optBWsskdeGaussian}
h_i^* = \left( \frac{1}{2 B^2|\bm{\mathcal{X}}_i| \sqrt{\pi}\Upsilon(f^{''}(x|\bm{\mathcal{X}}_i)) \mathbb{P}(\bm{\mathcal{X}}_i)^2}\right)^{\frac{1}{5}}.
\end{IEEEeqnarray}
\end{theorem}
\begin{proof}
See Appendix \ref{app:est}.
\end{proof}

\subsection{Optimal Bandwidth Computation}
The computation of (\ref{eq:optBWsskdeGaussian}) requires the knowledge of the true \ac{pdf}, as for (\ref{eq:optBWkde}). Therefore, to compute the optimal bandwidth, we resort to an iterative algorithm.
The optimal bandwidth for \ac{sbsse} has been proven to depend on both the probability $\mathbb{P}(\bm{\mathcal{X}}_i)$ and the second derivative of the \ac{pdf}, $f^{''}(\bm{x}|\bm{\mathcal{X}}_i)$. However, again, the true \ac{pdf} is not known. Therefore, we propose a heuristic algorithm, splitting the \ac{sbsse} design into the design of one \ac{kde} for each subset $i=1, \ldots, B$, obtaining a first estimate of the bandwidths $h_i$,  then refined iteratively. 

In particular, we note that (\ref{eq:optBWsskdeGaussian}) and (\ref{eq:optBWkde}) are similar, except for a multiplicative factor given by $\left(N/(B^2|\bm{\mathcal{X}}_i|\mathbb{P}(\bm{\mathcal{X}}_i)^2)\right)^{1/5}$. Therefore, we exploit the result obtained via \ac{kde} to obtain the optimal bandwidth for \ac{sbsse}.

Once the $i^{\rm th}$ bandwidth has been obtained from \ac{kde}, a first estimate of the probability $\hat{\mathbb{P}}(\bm{\mathcal{X}}_i)$ is obtained by integration over the $i^{\rm th}$ subset, i.e.,
\begin{equation}\label{eq:pSec}
\hat{\mathbb{P}}(\bm{\mathcal{X}}_i) = \int_{\bm{x} \in \bm{\mathcal{X}}_i} \hat{f}(\bm{x}|\bm{\mathcal{X}}_i)d\bm{x}.
\end{equation}
The bandwidth values are then updated according to the estimate (\ref{eq:pSec}), and the procedure is repeated until convergence. Given the \ac{pdf} estimates $\tilde{f}_{q}$ and $\tilde{f}_{q-1}$ respectively at iteration $q$ and $q-1$, the algorithm reaches convergence when the two \acp{pdf} are similar in terms of their Kullback-Leibler divergence \cite{cover1999elements}, i.e., when for a suitable parameter $\epsilon$
\begin{equation}
D_{\rm KL}(\tilde{f}_{q}||\tilde{f}_{q-1})=\sum_{\bm{x} \in \bm{\mathcal{X}}} \tilde{f}_{q}(\bm{x})\log_2 \frac{\tilde{f}_{q}(\bm{x})}{\tilde{f}_{q-1}(\bm{x})} \leq \epsilon,
\end{equation}
where $\log_2(\cdot)$ represents the base-$2$ logarithm.

The algorithm steps are presented in Algorithm \ref{alg:optBWsskde}.
\begin{algorithm}[t]\label{alg:optBWsskde} \small
 \KwData{$\bm{\mathcal{X}}$, $B$, $\bm{\mathcal{X}}_i \, \forall \, i=1,\cdots,B$, $\epsilon$}
 \KwResult{ $\hat{f}$}
 initialize $\tilde{f}_{0} = 0$\;
 compute $h_i^{(0)}$  \eqref{eq:optBWkde} via algorithm in \cite{botev2010}\;
 compute $\hat{f}(x|\bm{\mathcal{X}}_i)$ via \ac{kde} algorithm in \cite{botev2010}\;
 compute $\hat{\mathbb{P}}(\bm{\mathcal{X}}_i)$ via (\ref{eq:pSec}) $\forall i=1,\cdots B$\;
       $h_i^{(1)} = h_i^{(0)} \left(N/(B^2|\bm{\mathcal{X}}_i|\mathbb{P}(\bm{\mathcal{X}}_i)^2)\right)^{1/5}$, \, $\forall i=1,\cdots, B$\;
       compute $\tilde{f}_{1}$ using $h_i^{(1)}$\;
       q = 1\;
 \While{$D_{\rm KL}(\tilde{f}_{(q)}||\tilde{f}_{(q-1)}) > \epsilon$}{
       $q = q + 1$\;
       compute $\hat{f}(x|\bm{\mathcal{X}}_i)$ via KDE using $h_i^{(q-1)}$\;
 compute $\hat{\mathbb{P}}(\bm{\mathcal{X}}_i)$ via (\ref{eq:pSec}) $\forall i=1,\cdots B$\;
       $h_i^{(q)} = h_i^{(q-1)} \left(N/(B^2|\bm{\mathcal{X}}_i| \mathbb{P}(\bm{\mathcal{X}}_i)^2)\right)^{1/5}$, \, $\forall i=1,\cdots, B$\;
       compute $\tilde{f}_{(q)}$ using $h_i^{(q)}$\;
    }

 \caption{Bandwidth computation algorithm for \ac{sbsse}}
\end{algorithm}

\subsection{Low-complexity \ac{sbsse}}\label{sec:lowComp}
The proposed \ac{sbsse} algorithm has the drawback of a high computational complexity than \ac{kde}, as it requires $B$ \acp{kde} for each iteration. In order to reduce its complexity, we replace the bandwidth estimation algorithms with the estimation of a covariance matrix. In particular, since the measured \acp{ipv} are correlated due to the round robin scheduler, we compute the bandwidth as the correlation matrix of the \ac{ipv}, thus capturing time correlation. We consider the multidimensional Gaussian kernel, 
\begin{equation}
\begin{split}
& \Xi \left(\frac{\bm{s}_n-\bm{x}}{\bm{H}_i} \right) = \\
& \frac{1}{\sqrt{(2\pi)^D {\rm det}\bm{H}_i}}\exp\left(-\frac{(\bm{s}_n-\bm{x})^T\bm{H}_i^{-1}(\bm{s}_n-\bm{x})}{2} \right),
\end{split}
\end{equation}
where $\bm{H}_i = \mathbb{E}(\bm{s}_n \bm{s}_n^H)$ is the covariance matrix of the \ac{ipv} sequence in $i^{\rm th}$ subset and ${\rm det}\bm{H}_i$ its determinant. For each subset of $\bm{\mathcal{X}}$, we estimate the $D$-dimensional sample covariance matrix. We denote as $\bm{A}_{\ell}$ the $N_{\ell} \times D$ matrix whose rows are given by \ac{ipv} sequence of length $D$ jointly belonging to the $\ell^{\rm th}$ element of $\bm{\mathcal{X}}$. Denoting as $\mu_d$ the sample mean value of the $d$-th column, the $(q,j)$ element of the $i^{\rm th}$ sample covariance matrix is given by 
\begin{equation}\label{eq:bwVar}
[\bm{H}_i]_{q,j} = \frac{1}{N_{\ell}-1}\sum_{n=1}^{N_{\ell}}(\bm{A}_{.,q}-\mu_q)^*(\bm{A}_{.,j}-\mu_j).
\end{equation}
The resulting \ac{sbsse} density estimator is obtained as
\begin{equation}
    \tilde{f}(\bm{x}) = \frac{1}{B}\sum_{i=1}^{B}\frac{1}{|\bm{\mathcal{X}}_i|{\rm det}(\bm{H}_i)}\sum_{\bm{s} \in \bm{\mathcal{X}}_i}\Xi\left(\frac{\bm{s}-\bm{x}}{\bm{H}_i} \right).
\end{equation}

By replacing the iterative algorithm with an estimate of the sample covariance matrix, we reduced the computational complexity to a linear function of the number of measured \acp{ipv}. 
We henceforth denote the \ac{sbsse} with bandwidth given by (\ref{eq:bwVar}) as \ac{lcsb}.

\section{Computational Complexity} \label{sec:compcomp}
We here consider the computational complexity in terms of total number of addition and multiplication operations. The implementation of the \ac{kde} and of its optimal bandwidth in \cite{botev2010} requires the computation of the \ac{dct} of the data in the training set. The computational complexity of a two-dimensional \ac{dct} is given by
\cite{plonka2005fast}
\begin{equation}
\mathcal{C}_{\rm DCT}(\mu) = \frac{7}{3} \mu \beta - \frac{20}{9}\mu +\frac{2}{9}(-1)^\beta + 2,
\end{equation}
where $\mu$ is the length of the \ac{dct} and $\beta=\log_2\mu$.
Successively, it requires to find the root of the obtained function, with a total number of operations given by \cite [Th. 2.1]{burden1985numerical} 
\begin{equation}
\mathcal{C}_{\rm root} = 2 \log_2\frac{\eta_0}{\eta},
\end{equation}
where $\eta_0$ is the difference between the maximum and minimum value in the training set, and $\eta$ is the error tolerance.
The overall complexity is hence given by
\begin{equation}
\mathcal{C}_{KDE}(\mu) = \mathcal{C}_{\rm DCT}(\mu) + \mathcal{C}_{\rm root}.
\end{equation}

The \ac{sbsse} algorithm requires an initial \ac{kde} estimate for each subset. Then, it requires, for each iteration until convergence, the computation of a summation over the number of samples in each subset and 4 multiplications for each subset. The overall complexity is hence given by
\begin{equation}
\mathcal{C}_{\ac{sbsse}} = B \mathcal{C}_{KDE}(\mu / B) + n_{\rm it}4B\sum_{i=1}^B |\bm{\mathcal{X}}_i|,
\end{equation}
where $n_{it}$ denotes the number of iterations for convergence.

The \ac{lcsb} requires the estimation of the set variance for each subset, which is linear in the cardinality of the considered subset. Therefore, the overall complexity can be expressed as
\begin{equation}
\mathcal{C}_{\ac{lcsb}} = \sum_{i=1}^B |\bm{\mathcal{X}}_i|.
\end{equation}

\begin{figure}[t]
\centering
\includegraphics[width = 8.8cm]{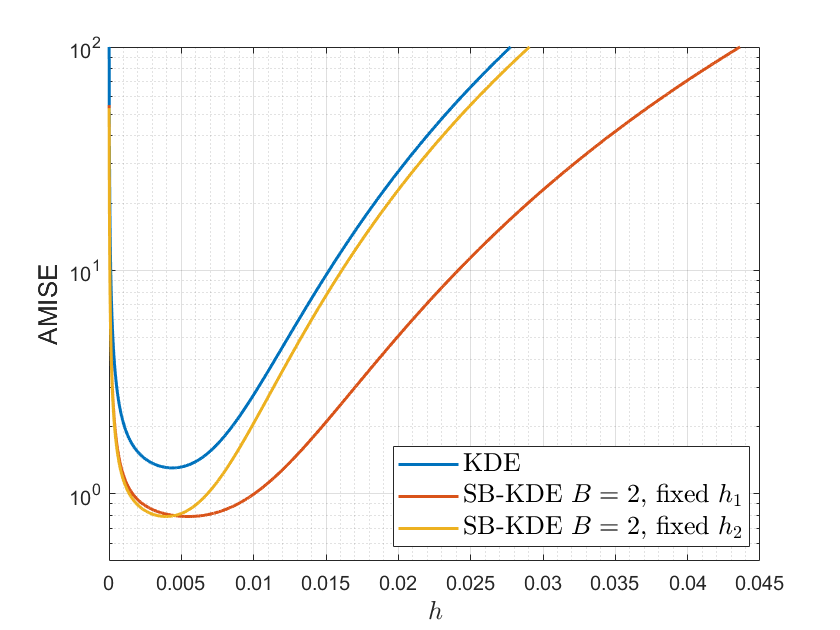}
\caption{\ac{amise} vs. bandwidth for \ac{kde} and \ac{sbsse}. Notice that, being \ac{amise} a function of $B=2$ bandwidths for \ac{sbsse}, curves are obtained by fixing one bandwidth value and varying the other. We assumed $S_{\rm max} = 100$, and true \ac{pdf} with $\bm{h}=[1.5, 2.7, 1.2]$.}
\label{fig:amise}
\end{figure}

\begin{figure*}[t]
	\subfigure[center][\ac{lcsb}V]
	{\includegraphics[width = 8.8cm]{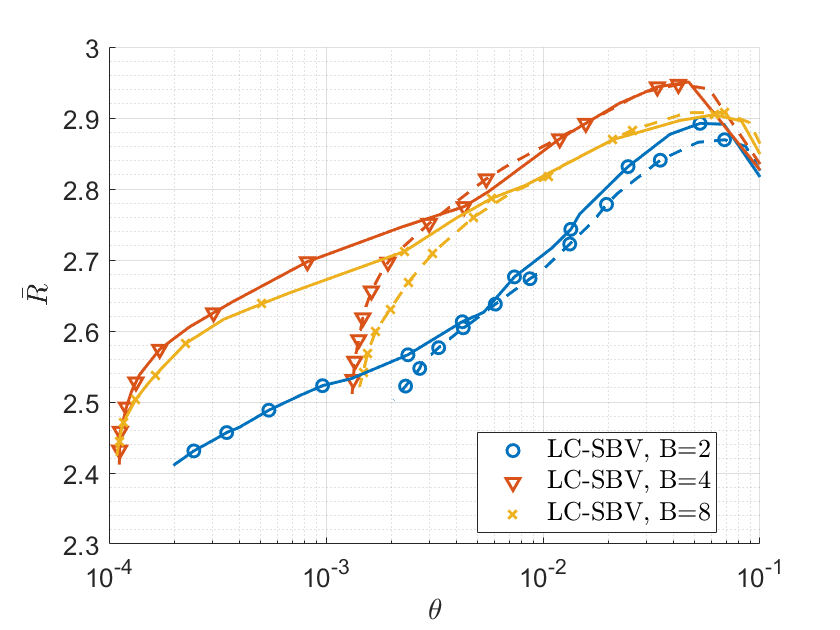}
		\label{fig:B_Rice}}
	\subfigure[center][\ac{lcsb}N]{
		\includegraphics[width = 8.8cm]{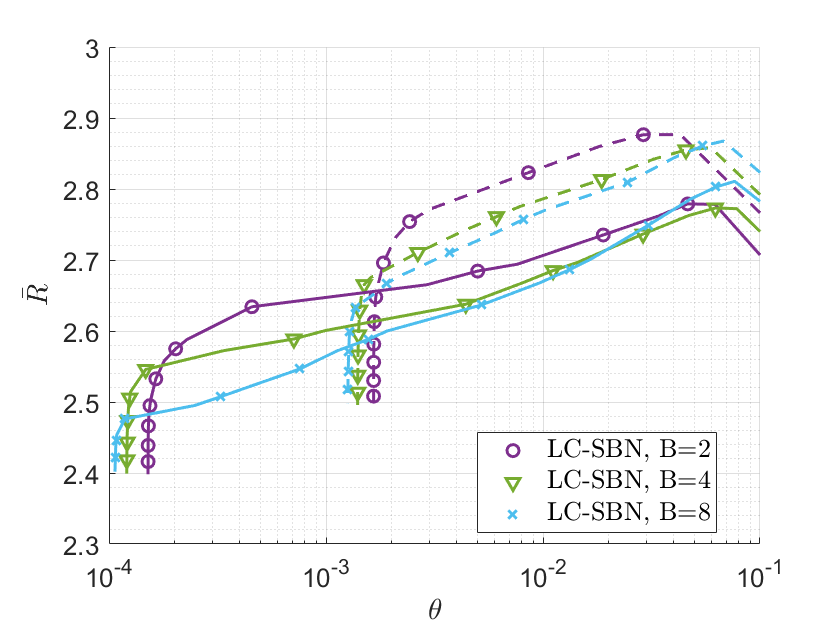}    
		\label{fig:H_Rice}}
	\caption{Average \ac{se} vs. $\theta$ for \ac{lcsb}V (left) and \ac{lcsb}N (right). In both figures same marker denotes the same number of subsets. Results are reported for a training set of $L=10^3$ samples (dashed lines) and a training set of $L=10^4$ samples (solid lines).}
	\label{fig:comparisonRice}
\end{figure*}

\section{Numerical Results}\label{sec:numRes}
\subsection{\ac{sbsse} Optimality}
In order to assess the performance of the \ac{sbsse} density estimator, we compare the \ac{amise} obtained for both \ac{kde} and \ac{sbsse} for a fixed and known \ac{pdf} $f(x)$ of a random variable. In particular, we assume that the true \ac{pdf} is given by (\ref{eq:sbkde}) for $B=3$ and $\bm{h}=[1.5, 2.7, 1.2]$. This allows us to compute the \ac{pdf} $f(x)$ for each $x$ in a predefined dataset $\mathcal{S}$. We assume that $\mathcal{S} =\{0,1, \ldots, S_{\rm max}\}$, and we randomly create sets $\mathcal{X}_i$, $i = 1,\ldots,B$, such that $\bigcup\limits_{i=1}^{B}\mathcal{X}_i = \mathcal{S}$.

Fig. \ref{fig:amise} shows the \ac{amise} obtained with \ac{kde} as in (\ref{eq:amiseKde}) and \ac{sbsse} as in (\ref{eq:amiseMult}) with $B=2$. Notice that both estimates are approximations of the true \ac{pdf}, which is based on $3$ subsets. Furthermore, since the \ac{amise} for \ac{sbsse} is a function of two bandwidths, results are shown by fixing one bandwidth to the optimal value, and letting the other vary. By using multiple bandwidths, the \ac{amise} is reduced, validating our approach. Furthermore, we notice that the proposed solution is less sensitive to sub-optimal bandwidth values. In fact, we notice that near-to-optimal \ac{amise} values are obtained for a wider range of bandwidth values for \ac{sbsse} with rspect to \ac{kde}.

\subsection{Link Adaptation}\label{sec:la_nr}
In this work, we assume that the \ac{la} can match the \ac{bler} target if the predicted \ac{ipv} is larger than that experienced at transmission time. Therefore, we assume a correct transmission if the predicted \ac{ipv} is larger than the actual one at the successive time instant, whereas otherwise a failure occurs. This models a system where uncontrolled re-transmissions are to be avoided, for instance due to very strict latency constraints in \ac{urllc}. For each \ac{ipv} test sequence of $T$ \acp{tti}, we evaluate the reliability of the different solutions by counting the number of events in which the predicted \ac{ipv} is below the actual one. We hence define the reliability of the system (between $0$ and $1$) as
\begin{equation}
1-\theta = 1-\frac{\sum_{t=1}^{T-1} \chi(t)}{T-1},
\end{equation}
where $\chi(t)$ is the indicator function
\begin{equation}
\chi(t) =
\begin{cases}
1 \quad {\rm if} \, \hat{\phi}(t) < \phi(t); \\
0 \quad {\rm otherwise}.
\end{cases}
\end{equation}
Notice that, $\theta$ represents the unreliability. Since we are interested in high reliability, we aim at small $\theta$ values.

For each \ac{ue} we assume that, if a failure happens, the experienced \ac{se} is zero, due to the packet's transmission's failure. Therefore, considering short packets of size $M$, the instantaneous \ac{se} $R(t)$ at \ac{tti} $t$ is \cite{polyanskiy2010channel} 
\begin{equation}\label{eq:se}
R(t) \approx (1-\chi(t))\left(\log_2(1+\hat{\rho}(t))- \sqrt{\frac{1}{M}V(\hat{\rho}(t))}Q^{-1}(\varepsilon)\right) ,
\end{equation}
where $V(\hat{\rho}(t)) = \frac{\hat{\rho}(t)(2+\hat{\rho}(t))}{(1+\hat{\rho}(t))^2}$, where $Q$ is the Gaussian complementary \ac{cdf}, $Q^{-1}$ is its inverse, and $\hat{\rho}(t)$ is the predicted \ac{sinr}. We denote as $\bar{R}$ the average \ac{se} in time, and as $M=128$ the number of channel uses. Notice that we used $\approx$ in (\ref{eq:se}) as we neglect terms with $O(\log M/M)$.

For the \ac{mq}-based methods, we consider that prediction exploits the conditional \ac{pdf} (\ref{eq:condPDF}), where prediction is based on the previous measured \ac{ipv}, i.e., $N_{\rm prev}=1$.

We compare two different heuristic policies for creating \ac{lcsb} subsets. The first is based on the values assumed by the \acp{ipv} in the time series. In particular, assuming an \ac{ipv} sequence with minimum value $m$ and maximum value $M$, the $b^{\rm th}$ subset is given by $ \mathcal{X}^{(b)}_d = x_d, \, {\rm s.t.} \, x_d \geq (b-1)\frac{M-m}{B}+1 \, {\rm and} \, x_d < b\frac{M-m}{B}$. We denote this method by  \ac{lcsb}V. The latter policy is based on the number of different \acp{ipv} in the time series. In particular, we populate each subset with the same number of sample points. We denote this method as  \ac{lcsb}N.

\subsection{Baseline Algorithms}
In order to better assess the performance of the \ac{sbsse} and its low-complexity version, we compare the results with four state-of-the-art algorithms. Two baselines are given by the \ac{mq} method proposed in \cite{vtc}, based on the empirical \ac{cdf} (\ref{eq:empCDF}) (henceforth referred to as ECDF) and on \ac{kde}. The third method is the based on \ac{olla}, described in Section \ref{sec:sysMod}. Since \ac{olla} does not exploit any interference prediction method, for a fair comparison, we exploit the low-pass filtering of the \acp{ipv} proposed in \cite{pocovi2018}. We will hereafter denote this method as \ac{olla}-LPP. 
In particular, given the previously predicted \ac{ipv} $\hat{\phi}(t)$ and the current measured \ac{ipv} $\phi(t)$, the \ac{ipv} to be used for the next transmission is given by \cite{pocovi2018}
\begin{equation}
\hat{\phi}(t+1) = \alpha \phi(t) + (1-\alpha)\hat{\phi}(t-1),
\end{equation}
where $\alpha$ is a constant real value, which is usually small. The predicted value is then used to compute the \ac{sinr} $\rho_{OLLA}^{(t+1)}$.
The last baseline method assumes the \acp{ipv}' distribution to be log-normal. Its first and second (order) moments are estimated as a log-normal random variable \cite{bb}. We propose to set the first and second moments of the \ac{pdf} as the mean $\mu_{\bm{\mathcal{X}}}$ and variance $\sigma^2_{\bm{\mathcal{X}}}$ of the training set given by $\bm{\mathcal{X}}$. The predicted \ac{ipv} is hence given by 
\begin{equation}
\hat{\phi}(t+1) = \exp\left(\Theta^{-1}\left(2\varepsilon-1 \right)\sqrt{2}\sigma_{\bm{\mathcal{X}}}+\mu_{\bm{\mathcal{X}}}\right),
\end{equation}
where $\Theta(x)$ represents the error function
\begin{equation}
\Theta(x) = \frac{1}{\sqrt{\pi}}\int_{-x}^x e^{-t^2}dt.
\end{equation}

\subsection{Rice Channel Model}
We here consider a scenario with $N=9$ square cells, with each \ac{gnb} located at the center of is at a distance of $200$~m from neighboring \acp{gnb}. Each \ac{gnb} is equipped with $N_a=16$ antennas linearly spaced by $d=\lambda/2$. The \acp{ipv} are measured in the central cell, and the number of \acp{ue} in each surrounding cell is given by the realization of a uniform random variable in range $[2, 8]$. We consider a noise power of $\sigma^2 = -101$~dBm, a transmitted power at each \ac{gnb} of $P=46$~dBm, Rice $K$ factor $\Psi=10$~dB, path loss exponent $\nu=3.5$ and a cell edge \ac{snr} without interference of $20$~dB. This corresponds to a typical highly interference limited scenario in practical deployments \cite{138901}.

Fig. \ref{fig:comparisonRice} shows the average \ac{se} vs. $\theta$ for the  \ac{lcsb}V (left) and  \ac{lcsb}N (right). Results have been obtained by fixing the number $L$ of training samples used for \ac{pdf} estimation and changing the parameter $\varepsilon$, yielding different values of $\theta$. Figures show both the impact of increasing $L$ and the effect of the number of subsets $B$. About the sensitivity with respect to $L$, we denote with dashed line the results obtained estimating the \ac{pdf} with $L=10^3$ samples and with solid line the results obtained estimating the \ac{pdf} with $L=10^4$ samples. Moreover, for each curve in each figure, a higher reliability, i.e., la smaller $\theta$, is obtained by decreasing $\varepsilon$. We notice that a larger $L$ leads to a higher reliability with both policies, since having a larger number of training samples yields a higher precision in the estimated \ac{pdf}. We also notice that for  \ac{lcsb}N a lower $L$ leads to higher \ac{se} for high $\theta$. Indeed, a lower precision in the \ac{pdf} estimation leads to a less conservative behavior, with higher rates and a lower reliability. We now consider the effect of the number of subsets. We notice that more subsets entail a lower $\theta$ and hence a higher reliability for both policies, whereas in terms of rate results do not show the same behavior. In fact, for  \ac{lcsb}V the best trade-off between \ac{se} and reliability is obtained when considering $B=4$ subsets, whereas for  \ac{lcsb}N is obtained with $B=2$ subsets. We can hence identify the number of subsets as a hyper-parameter, to be optimized based on collected data. Comparing the two policies, we notice that \ac{lcsb}V attains higher \ac{se} as the reliability increases. In the following, we choose  \ac{lcsb}V $B=4$ as the policy to be compared with baseline algorithms, as it achieves the best performance both in terms of average \ac{se} and reliability.
\begin{figure}[t]
\centering
\includegraphics[width=8.8cm]{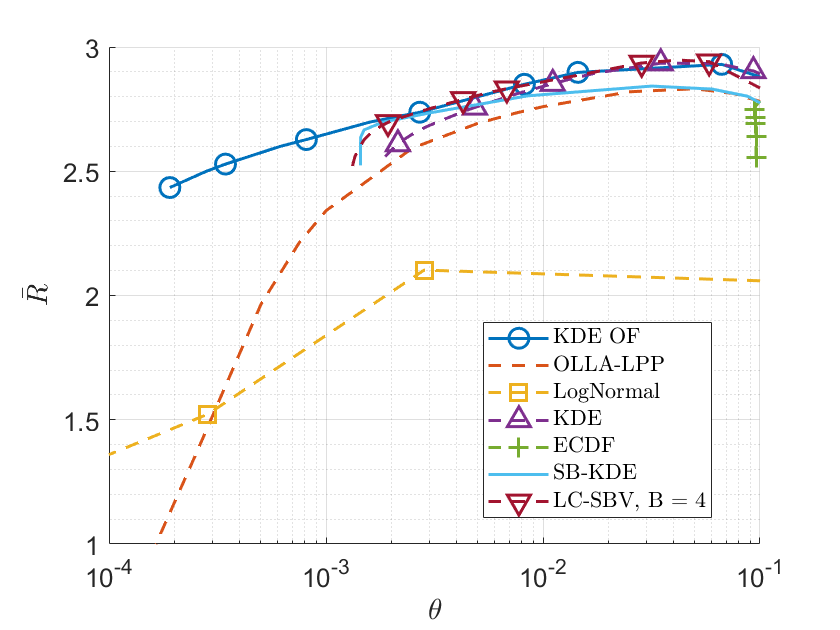}
\caption{Average \ac{se} vs. $\theta$ for the best performing policy chosen from Fig. \ref{fig:comparisonRice} and the baseline methods. Estimation performed with $L=10^3$ samples.}
\label{fig:compRice100}
\end{figure}

Fig. \ref{fig:compRice100} shows the average \ac{se} vs. $\theta$ for the best policy chosen from Fig. \ref{fig:comparisonRice} and the baseline methods, when considering $L=10^3$ training samples. For \ac{kde} we show both the performance obtained with $L=10^3$ training samples and the optimal results obtained by over-fitting, i.e., using all samples in the dataset for training. The ECDF method based on (\ref{eq:empCDF}) attains a poor approximation of the \ac{cdf}, since not enough points are available to match the strict reliability targets. The comparison with the other methods motivates the need for kernel-based density estimators. About \ac{olla}-LPP we notice that, although it works well with high $\theta$ values, the attained \ac{se} rapidly degrades for decreasing $\theta$, due to the highly conservative behavior of \ac{olla} which favors reliability to \ac{se}. Considering a \ac{urllc} scenario targeting  high reliability, and hence small $\theta$, we notice that \ac{olla}-LPP is not able to guarantee both high reliability and high \ac{se}, therefore not being a suitable solution if the load offered by \ac{urllc} becomes relevant. A similar behavior is obtained with the LogNormal approximation, where \ac{se} rapidly degrades with increasing reliability values. However, differently from \ac{olla}-LPP, the \ac{se} has a slower decrease and does not tend to zero for $\theta < 10^{-4}$. Furthermore, we notice that, compared to the other \ac{pdf}-based methods, LogNormal reaches smaller values of $\theta$. Indeed, this method relies on a closed-form equation of the \ac{ipv} \ac{pdf} and does not suffer for an insufficient number of training data, being able to reach infinite precision. On the other hand, we notice that the LogNormal approximation attains a higher \ac{se} than \ac{olla}-LPP for $\theta < 2 \cdot 10^{-4}$. The \ac{se} for the proposed \ac{sbsse} degrades as $\theta$ decreases. However, it attains a higher \ac{se} when it's able to match the desired reliability when compared to LogNormal and \ac{olla}-LPP, whereas, when compared to \ac{kde}, it attains higher \ac{se} only for $\theta \leq 5 \cdot 10^{-3}$.   Furthermore, compared to \ac{kde}, it also to attains lower $\theta$, for $\theta \leq 2 \cdot 10^{-3}$, confirming that a subset-based approach is advantageous over the plain \ac{kde}. Comparing the low-complexity  \ac{lcsb}V with all other approaches, we notice that it attains the highest \ac{se} for all $\theta \geq 2 \cdot 10^{-3}$. However, LC-SBV is limited by the demanding amount of data needed to estimate the \ac{pdf}.
\begin{figure} 
\centering
\includegraphics[width=8.8cm]{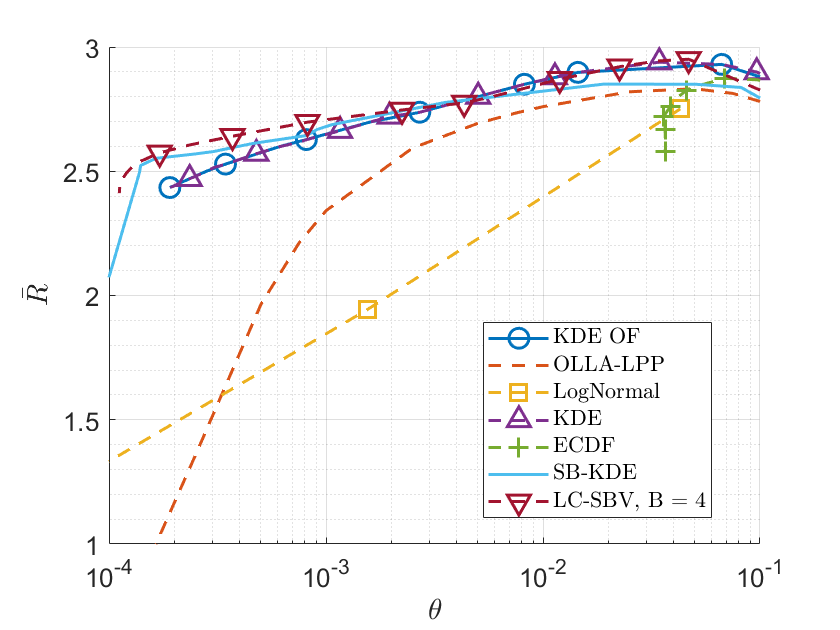}
\caption{Average \ac{se} vs. $\theta$ for the best performing policy chosen from Fig. \ref{fig:comparisonRice} and the baseline methods. Estimation performed with $L=10^4$ samples.}
\label{fig:compRice}
\end{figure}
\begin{figure*} 
\centering
	\subfigure[center][\ac{lcsb}V]
	{\includegraphics[width = 8cm]{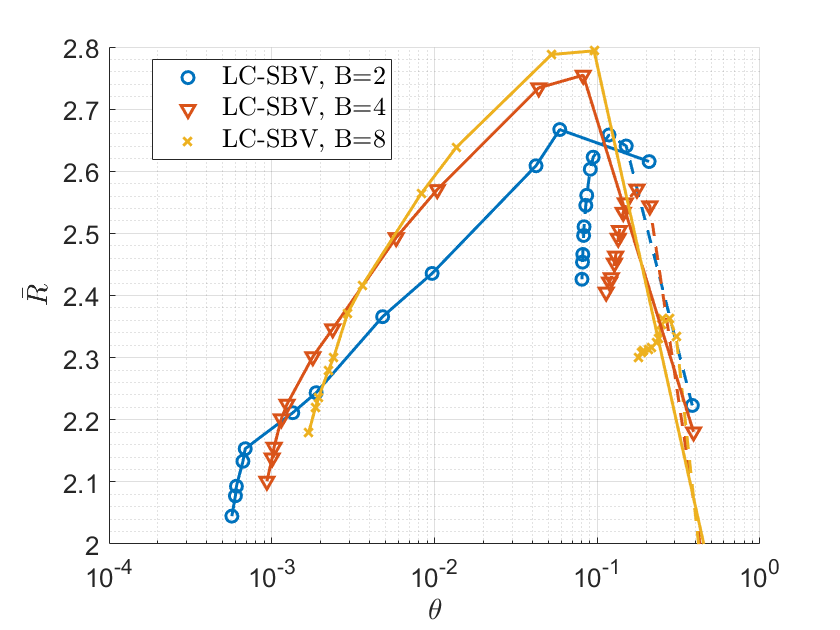}}
	\subfigure[center][\ac{lcsb}N]{
		\includegraphics[width = 8cm]{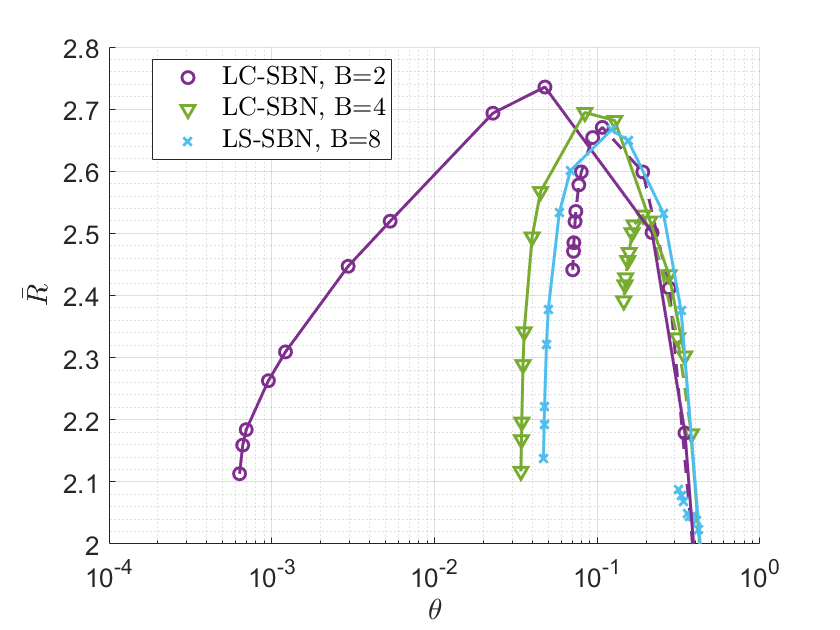} }   
	\caption{\ac{se} vs. $\theta$ for \ac{lcsb}V (left) and \ac{lcsb}N (right) obtained with the \ac{3gpp} channel model. In both figures same marker shape denotes the same number of subsets. Results are reported for a training set of $L=10^2$ samples (dashed lines) and a training set of $L=10^4$ samples (solid lines).}
	\label{fig:comparison3GPP}
\end{figure*}
\begin{figure} 
\centering
\includegraphics[width=8.8cm]{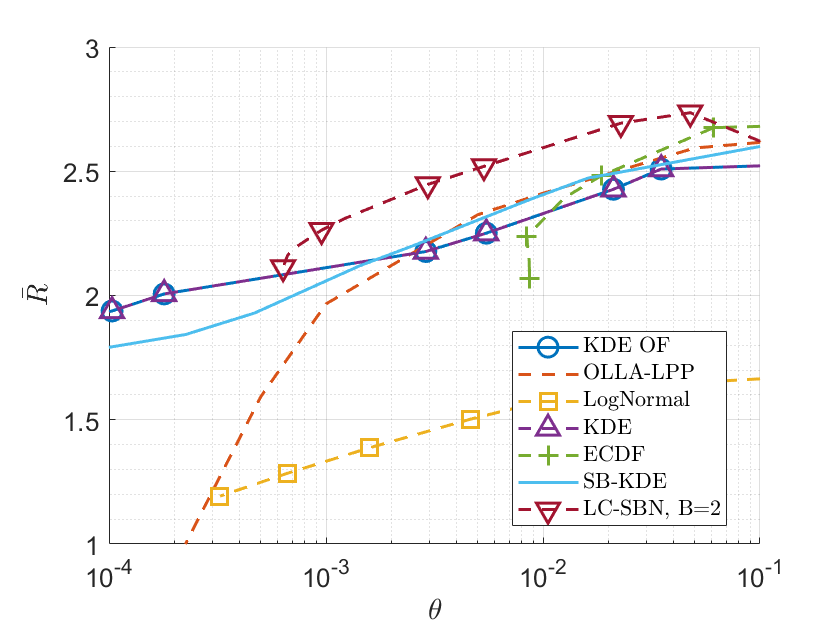}
\caption{Average \ac{se} vs. $\theta$ for the best performing policy chosen from Fig. \ref{fig:comparison3GPP} and the baseline methods. Results obtained with $L=10^4$ training samples.}
\label{fig:comp3gpp}
\end{figure}
Fig. \ref{fig:compRice} shows the average \ac{se} vs $\theta$, for the best performing policy chosen from Fig. \ref{fig:comparisonRice} and the baseline methods, when considering $L=10^4$ training samples. Note that both \ac{kde} approaches have similar performance, i.e., convergence is already obtained with $10^4$ training samples. For the ECDF method, a larger training set does not improve its performance. About \ac{sbsse}, we notice a \ac{se} degradation smaller than $0.2$ bit/s/Hz, up to $\theta=0.5 \cdot 10^{-4}$, then \ac{se} rapidly decreases, still being higher than that of \ac{olla}-LPP and that of Log-normal approximation. Furthermore, we also notice that the subset-based approaches attain smaller $\theta$ values than \ac{kde}, and \ac{lcsb}V is the best performing method down to $\theta=10^{-4}$.

\subsection{3GPP Channel Model}
In order to test the proposed algorithm in a more realistic scenario, we consider in this section a three-dimensional spatial 3D Urban Micro (UMi) channel, calibrated with the results obtained by \ac{3gpp} \cite{138901}. In detail, we consider $C=21$ cells, organized in $7$ sites, each with $120$ degrees subsets per cell, inter-site distance of $200$ m, and wraparound. An average of $5$ \acp{ue} is deployed per cell, which move with a speed of $3$ km/h. Each \ac{gnb} is equipped with $64$ antennas, organized in a uniform planar array, with $8$ rows, $4$ columns and with cross-polarized antenna elements. The \acp{gnb} serve \acp{ue} at a carrier frequency of $3.7$ GHz on a system bandwidth of $10$ MHz and using a transmit power of $41$ dBm. We follow the scenario in \cite{138901}, where the reader can find more details.

Results have been obtained by fixing the number $L$ of training samples for \ac{pdf} estimation and varying $\varepsilon$, allowing to obtain different $\theta$ values. As for the Rice channel model, we show both the effect of increasing $L$ and varying the number $B$ of subsets. 

Fig. \ref{fig:comparison3GPP} shows the \ac{se} vs. $\theta$ for the two considered policies for the \ac{lcsb} algorithm. Dashed lines refer to training sequences of $L=10^2$ samples, while solid lines to training sequences of $10^4$ samples. We notice that, if the training set does not have enough samples, both methods are unable to attain small $\theta$ values with a poor reliability. About \ac{lcsb}V, we notice the trade-off between \ac{se} and reliability when considering the number of subsets $B$ needed to estimate the \ac{pdf}. In fact, on one hand a larger number of subsets yields higher \ac{se} values, on the other hand a smaller number of subsets means we  reach smaller $\theta$ values. Hence, this parameter will be set according to the specific users' need. About \ac{lcsb}N, we instead notice that increasing the number of subsets does not improve performance. However, the \ac{lcsb}N with $B=2$ is the best performing method among all policies and number of subsets, and is hence compared with the baseline methods.

Fig. \ref{fig:comp3gpp} shows the average \ac{se} vs. $\theta$ for all the baseline methods, the proposed \ac{sbsse} and \ac{lcsb}N with $B=2$ subsets. Results are obtained with $L=10^4$ training samples. As for Fig. \ref{fig:compRice}, both \ac{olla}-LP and the LogNormal approximation attain decreasing \ac{se} values for decreasing $\theta$.  Regarding \ac{kde}, performance obtained considering $10^4$ training samples is equal to those obtained considering the full dataset, meaning that convergence has been reached.  However we recall that, although these are the best results obtainable with \ac{kde}, they do not represent the overall optimum, as the estimated \ac{pdf} may be inaccurate. In fact, with ECDF we attain higher \acp{se} for high $\theta$ values. However, \ac{kde} can reach $\theta=10^{-4}$, showing that kernel methods   provide good performance for low reliability target. About the proposed \ac{sbsse}, it achieves higher \ac{se} values compared to those of \ac{kde} for $\theta > 2 \cdot 10^{-3}$, whereas \ac{lcsb}N attains higher \ac{se} values for all $\theta \geq 6 \cdot 10^{-4}$, as \ac{lcsb}N can not attain lower $\theta$ values considering $L=10^4$ training samples.

\begin{figure} 
\centering
\includegraphics[width=8.8cm]{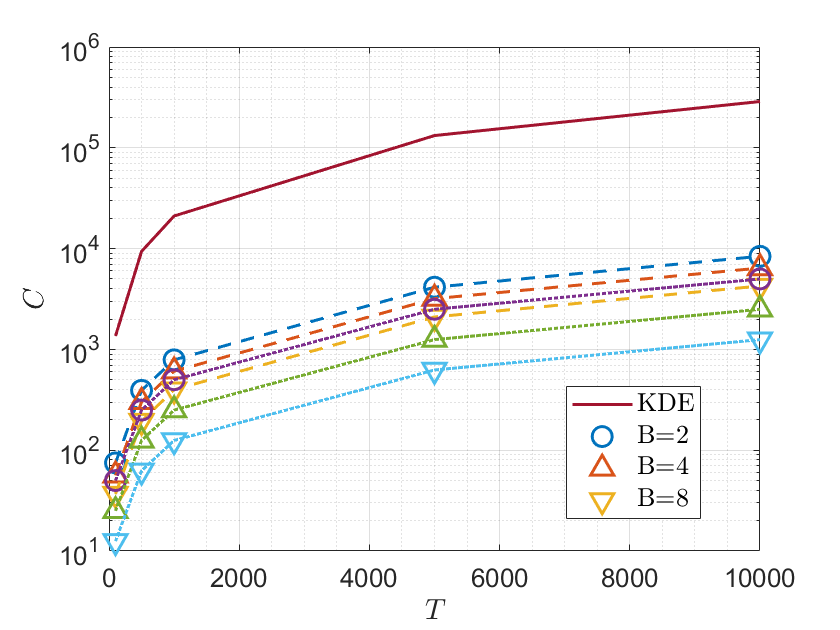}
\caption{Computational complexity vs. number of training data for the kernel based density estimators. Dashed curves show results for \ac{lcsb}V, whereas dotted curves show results for \ac{lcsb}N.}
\label{fig:ccRice}
\end{figure}

Fig. \ref{fig:ccRice} shows the computational complexity in terms of number of additions and multiplications vs. the number of training samples $L$ for the \ac{kde} and the two policies for the low-complexity  \ac{lcsb}. The best performing method is the \ac{lcsb}N, due to the fact that it linearly depends on the number of samples per subset, which decreases as the number of subset increases.

\section{Conclusions}\label{sec:conclusion}
In this paper we considered the problem of predicting interference power to enable efficient link adaptation for \ac{urllc}. We  considered variable bandwidth kernel density estimators. We first derived the optimal bandwidth for the \ac{sbsse}, and based on the optimal solution for \ac{kde}, we proposed a heuristic algorithm to estimate a \ac{pdf} based on the optimal bandwidths. Motivated by the considerable computational complexity of the proposed solution, we then proposed a low-complexity version of the \ac{sbsse}, namely the \ac{lcsb}. By means of extensive simulations in cellular networks, considering realistic \ac{3gpp} 3D UMi channel models, we showed through numerical evaluations that the proposed solutions  attain at the same time higher \ac{se} a better and matching of the reliability targets than state-of-the-art solutions. 
By jointly looking at results in Fig.s \ref{fig:compRice},\ref{fig:comp3gpp} and \ref{fig:ccRice}, we can conclude that the proposed \ac{lcsb} method attains the best \ac{se} with the lower computational complexity and achieves extremely high reliability targets. Therefore, it is the best investigated algorithm for \ac{urllc}, being hence an effective approach to \ac{la}.

\begin{appendices}
\section{Derivation of the Optimal Bandwidth for \ac{sbsse}}\label{app:est}
We henceforth consider operations between vector and scalar as element-wise. We also recall that the kernel function of a vector is a scalar value. We assume that $K(\cdot)$ is a kernel function as defined in \cite{parzen1962}, therefore $\int K(u) du =1$ and that $\int uK(u) du = 0$.

\subsection{Bias}
Define the local KDE as
\begin{equation}
    \hat{f}(\bm{x}) = \frac{1}{B}\sum_{i=1}^B\frac{1}{|\bm{\mathcal{X}}_i|}\sum_{\bm{z} \in \bm{\mathcal{X}}_i}\frac{1}{h_i}K\left( \frac{\bm{z}-\bm{x}}{h_i}\right).
\end{equation}
The mean value of the kernel function can be computed as
\begin{equation}\label{eq:avgKer}
\begin{split}
    & E\left[\frac{1}{h_i}K\left( \frac{\bm{z}-\bm{x}}{h_i}\right)\right] =  
     \int_{-\infty}^\infty\frac{1}{h_i}K\left( \frac{\bm{\zeta}-\bm{x}}{h_i}\right) f(\bm{\zeta}) d\bm{\zeta}. \\ 
    \end{split}
\end{equation}
From the total probability law we have
\begin{equation}
    f(\bm{\zeta}) = \sum_{\ell=1}^B f(\bm{\zeta}|\bm{\mathcal{X}}_{\ell})\mathbb{P}(\bm{\mathcal{X}}_{\ell}),
\end{equation}
and by substitution in (\ref{eq:avgKer}) we obtain
\begin{equation}\label{eq:avgKer2}
\begin{split}
    & E\left[\frac{1}{h_i}K\left( \frac{\bm{z}-\bm{x}}{h_i}\right)\right]  =  \\ 
    & \int_{-\infty}^\infty\frac{1}{h_i}K\left( \frac{\bm{\zeta}-\bm{x}}{h_i}\right) \sum_{\ell=1}^B f(\bm{\zeta}|\bm{\mathcal{X}}_{\ell})\mathbb{P}(\bm{\mathcal{X}}_{\ell}) d\bm{\zeta}. \\ 
\end{split}
\end{equation}
By performing the change of variables $\bm{u}=(\bm{z}-\bm{x})/h_i$ we obtain
\begin{equation}\label{eq:meanV}
\begin{split}
   & E\left[\frac{1}{h_i}K\left( \frac{\bm{z}-\bm{x}}{h_i}\right)\right]  = \\
   & \int_{-\infty}^\infty K\left( \bm{u}\right) \sum_{\ell=1}^B f(\bm{u}h_i+\bm{x}|\bm{\mathcal{X}}_{\ell})\mathbb{P}(\bm{\mathcal{X}}_{\ell}) d\bm{u}. 
\end{split}
\end{equation}
Let us consider the second order Taylor series expansion of $f(\bm{u}h_i+x|\bm{\mathcal{X}}_i)$, obtaining
\begin{equation}\label{eq:Taylor}
\begin{split}
  & f(\bm{u}h_i+\bm{x}|\bm{\mathcal{X}}_i) \approx f(\bm{x}|\bm{\mathcal{X}}_i) + f^{(1)}(\bm{x}|\bm{\mathcal{X}}_i)h_i\bm{u} + \\
  & \frac{1}{2}f^{''}(\bm{x}|\bm{\mathcal{X}}_i)h_i^2\bm{u}^2 + o(h_i^2).
\end{split}
\end{equation}
By substitution of (\ref{eq:Taylor}) in (\ref{eq:meanV}) we obtain
\begin{IEEEeqnarray}{lCr}
  & \sum_{\ell=1}^B\int_{-\infty}^\infty K\left( \bm{u}\right) f(\bm{u}h_i+\bm{x}|\bm{\mathcal{X}}_i)\mathbb{P}(\bm{\mathcal{X}}_{\ell}) d\bm{u}  =\\
  &\sum_{\ell=1}^B\int_{-\infty}^\infty K\left( \bm{u}\right) \big( f(\bm{x}|\bm{\mathcal{X}}_{\ell}) + f^{(1)}(\bm{x}|\bm{\mathcal{X}}_{\ell})h_{\ell}\bm{u} + \\
  & \frac{1}{2}f^{''}(\bm{x}|\bm{\mathcal{X}}_{\ell})h_{\ell}^2\bm{u}^2\big)\mathbb{P}(\bm{\mathcal{X}}_{\ell}) d\bm{u}  = \nonumber \\
  & \sum_{\ell=1}^B\left(f(\bm{x}|\bm{\mathcal{X}}_{\ell}) +  \frac{1}{2}f^{''}(\bm{x}|\bm{\mathcal{X}}_{\ell})h_{\ell}^2\omega_2(K)\right)\mathbb{P}(\bm{\mathcal{X}}_{\ell}), \nonumber
\end{IEEEeqnarray}
where the second term involving the first derivative is zero as the mean of $K(\bm{u})$ is zero and where $\omega_2(K)$ is defined in (\ref{eq:omegaDef}).

Therefore 
\begin{IEEEeqnarray}{lCr}
  & E[\hat{f}(\bm{x})] = 
   E\left[\frac{1}{B}\sum_{i=1}^B\frac{1}{|\bm{\mathcal{X}}_i|}\sum_{\bm{z} \in \bm{\mathcal{X}}_i}\frac{1}{h_i}K\left( \frac{\bm{z}-\bm{x}}{h_i}\right)\right] \nonumber \\
  & = \frac{1}{B}\sum_{i=1}^B\frac{1}{|\bm{\mathcal{X}}_i|}\sum_{\bm{z} \in \bm{\mathcal{X}}_i}E\left[\frac{1}{h_i}K\left( \frac{\bm{z}-\bm{x}}{h_i}\right)\right] \nonumber \\
  & =  \frac{1}{B}\sum_{i=1}^B\frac{1}{|\bm{\mathcal{X}}_i|}\sum_{\bm{z} \in \bm{\mathcal{X}}_i,\ell=1}^B\left(f(\bm{x}|\bm{\mathcal{X}}_{\ell}) +  \frac{1}{2}f^{''}(\bm{x}|\bm{\mathcal{X}}_{\ell})h_{\ell}^2\omega_2(k)\right)\mathbb{P}(\bm{\mathcal{X}}_{\ell}) \nonumber\\
  & =  f(\bm{x}) + \frac{1}{B}\sum_{i=1}^B \left( \sum_{\ell=1}^B\frac{1}{2}f^{''}(\bm{x}|\bm{\mathcal{X}}_{\ell})h_{\ell}^2\omega_2(k)\mathbb{P}(\bm{\mathcal{X}}_{\ell})\right) \nonumber \\
   & =  f(\bm{x}) +  \sum_{\ell=1}^B\frac{1}{2}f^{''}(\bm{x}|\bm{\mathcal{X}}_{\ell})h_{\ell}^2\omega_2(k)\mathbb{P}(\bm{\mathcal{X}}_{\ell}). \nonumber 
\end{IEEEeqnarray}
Therefore, the bias is
\begin{IEEEeqnarray}{lCr}\label{appbias}
  & {\rm Bias}(\hat{f}(\bm{x})) = E[\hat{f}(\bm{x})] -f(\bm{x})\\
  & = \omega_2(k)\sum_{\ell=1}^B\frac{1}{2}f^{''}(\bm{x}|\bm{\mathcal{X}}_{\ell})h_{\ell}^2\mathbb{P}(\bm{\mathcal{X}}_{\ell}) + o(h_i^2). \nonumber
\end{IEEEeqnarray}

\subsection{Variance}
From the bias analysis we saw how the kernel function of a vector can be treated as a random variable. The variance of a \ac{rv} $x$ can be computed as
\begin{equation}
    {\rm Var}(x) = E[x^2] - E[x]^2.
\end{equation}
From the bias analysis, focusing on the first order Taylor approximation we have
\begin{IEEEeqnarray}{lCr}
    & E[\hat{f}(\bm{x})]  \approx f(\bm{x}) + o(1),
\end{IEEEeqnarray}
where the second term is $O\left(\frac{1}{N}\right)$, being $N$ the number of samples used for density estimation.
Then, following the approach used for the bias computation we have
\begin{IEEEeqnarray}{lCr}\label{eq:sqV}
   & E\left[\frac{1}{h_i}K\left( \frac{\bm{z}-\bm{x}}{h_i}\right)^2\right]  =  \\
   & \frac{1}{h_i}\int_{-\infty}^\infty K\left( \bm{u}\right)^2 \sum_{\ell=1}^B f(\bm{u}h_i+\bm{x}|\bm{\mathcal{X}}_{\ell})\mathbb{P}(\bm{\mathcal{X}}_{\ell}) d\bm{u} \\
   & = \int_{-\infty}^\infty K\left( \bm{u}\right)^2  f(\bm{u}h_i+\bm{x}) d\bm{u}, \nonumber
\end{IEEEeqnarray}
where the second equality comes from the fact that we considered a first order Taylor series and the total probability law.
Therefore, recalling that the kernel estimator is a linear estimator and that $K\left( \frac{\bm{z}-\bm{x}}{h_i}\right)$ is \ac{iid}
\begin{IEEEeqnarray}{lCr}\label{appvar}
  & {\rm Var}(\hat{f}(\bm{x}))  \\
  & = \frac{1}{B^2}\sum_{i=1}^B\frac{1}{|\bm{\mathcal{X}}_i|^2}\sum_{\bm{z} \in \bm{\mathcal{X}}_i}\frac{1}{h_i} \Upsilon(K)f(\bm{x}) + O\left(\frac{1}{N}\right), \nonumber \\
  & = \Upsilon(K)f(\bm{x})\frac{1}{B^2}\sum_{i=1}^B \frac{1}{|\bm{\mathcal{X}}_i|h_i}  + O\left(\frac{1}{N}\right), \nonumber
\end{IEEEeqnarray}
where $\Upsilon(K)$ is defined in (\ref{eq:rDef}).

\subsection{AMISE}
The MSE can be expressed as
\begin{IEEEeqnarray}{lCr}
    {\rm MSE} = {\rm Bias}(\hat{f}(\bm{x}))^2 + {\rm Var}(\hat{f}(\bm{x})).
\end{IEEEeqnarray}
The asymptotic mean squared error (AMSE) is obtained by the asymptotic derivation of bias and variance \eqref{appbias} and \eqref{appvar} as
\begin{equation}
\begin{split}
 & {\rm AMSE}
 =\left( \frac{1}{2}\omega_2(K)\sum_{i=1}^Bf^{''}(\bm{x}|\bm{\mathcal{X}}_i)h_i^2\mathbb{P}(\bm{\mathcal{X}}_i) \right)^2 \\
& + \Upsilon(K)f(\bm{x})\frac{1}{B^2}\sum_{i=1}^B\frac{1}{|\bm{\mathcal{X}}_i|h_i} . \nonumber
\end{split}
\end{equation}
By integrating the AMSE we obtain the AMISE
\begin{equation}
\begin{split}
& {\rm AMISE}  \\
& = \int_{-\infty}^\infty \left[\left( \frac{1}{2}\omega_2(K)\sum_{i=1}^Bf^{''}(\bm{x}|\bm{\mathcal{X}}_i)h_i^2\mathbb{P}(\bm{\mathcal{X}}_i) \right)^2 \right. + \\ & \left. \Upsilon(K)f(\bm{x})\frac{1}{B^2}\sum_{i=1}^B\frac{1}{|\bm{\mathcal{X}}_i|h_i} \right] d\bm{x} .
\end{split}
\end{equation}
Since $f(\bm{x}|\bm{\mathcal{X}}_i)=0 \, \forall \bm{x} \notin \bm{\mathcal{X}}_i$ we obtain
\begin{equation}
\begin{split}
& {\rm AMISE}  \\
& = \int_{-\infty}^\infty \left( \frac{1}{4}\omega_2(k)^2\sum_{i=1}^Bf^{''}(\bm{x}|\bm{\mathcal{X}}_i)^2h_i^4\mathbb{P}(\bm{\mathcal{X}}_i)^2 \right. \\
& \left.+ \Upsilon(K)f(\bm{x})\frac{1}{B^2}\sum_{i=1}^B\frac{1}{|\bm{\mathcal{X}}_i|h_i}  \right) d\bm{x} \nonumber \\
& = \sum_{i=1}^B\left(\frac{1}{4} \Upsilon(f^{''}(\bm{x}|\bm{\mathcal{X}}_i))h_i^4\mathbb{P}(\bm{\mathcal{X}}_i)^2\omega_2(k)^2 + \frac{\Upsilon(K)}{B^2|\bm{\mathcal{X}}_i| h_i} \right). \end{split}
\end{equation}

\subsection{Optimal bandwidth}
The optimal value $h_i^*$ can be computed as
\begin{IEEEeqnarray}{lCr}
\frac{\partial {\rm AMISE}}{\partial h_i}  &  = \Upsilon(f^{''}(\bm{x}|\bm{\mathcal{X}}_i))h_i^3\mathbb{P}(\bm{\mathcal{X}}_i)^2\omega_2(K)^2 - \frac{\Upsilon(K)}{B^2|\bm{\mathcal{X}}_i| h_i^2}  \nonumber \\
 & = 0, \nonumber
\end{IEEEeqnarray}
from which the AMISE optimal bandwidth value is given by
\begin{equation}
    h_i^* = \left( \frac{\Upsilon(K)}{B^2|\bm{\mathcal{X}}_i| \Upsilon(f^{''}(\bm{x}|\bm{\mathcal{X}}_i))\mathbb{P}(\bm{\mathcal{X}}_i)^2\rho_2(k)^2}\right)^{\frac{1}{5}}.
\end{equation}

Considering a Gaussian kernel we have 
\begin{IEEEeqnarray}{lCr}
&{\rm AMISE}  \\
& = \frac{1}{4}\sum_{i=1}^B\left( \Upsilon(f^{''}(\bm{x}|\bm{\mathcal{X}}_i))h_i^4\mathbb{P}(\bm{\mathcal{X}}_i)^2 + \frac{1}{B^2|\bm{\mathcal{X}}_i| 2\sqrt{\pi}h_i} \right), \nonumber
\end{IEEEeqnarray}
and therefore
\begin{IEEEeqnarray}{lCr}\label{eq:optMV}
h_i^* = \left( \frac{1}{2 B^2|\bm{\mathcal{X}}_i| \sqrt{\pi}\Upsilon(f^{''}(\bm{x}|\bm{\mathcal{X}}_i))\mathbb{P}(\bm{\mathcal{X}}_i)^2}\right)^{\frac{1}{5}}.
\end{IEEEeqnarray}
\end{appendices}
 
\bibliographystyle{IEEEtran}
\bibliography{bibliography}
\end{document}